\documentclass[times]{qjrms4}

\usepackage[
    colorlinks,
    bookmarksopen,
    bookmarksnumbered,
    citecolor=red,
    urlcolor=red
]{hyperref}

\newcommand\BibTeX{{\rmfamily B\kern-.05em \textsc{i\kern-.025em b}\kern-.08em
T\kern-.1667em\lower.7ex\hbox{E}\kern-.125emX}}

\usepackage{moreverb}
\usepackage{caption}
\usepackage{subcaption}
\usepackage{gensymb}

\def\volumeyear{2013}

\begin{document}

\title{Remote sensing of the temporal and spatial variability of atmospheric refractivity using ADS-B interferometry (ADSBi)}

\author{Ollie Lewis\affilnum{1}\corrauth, Chris Brunt \affil{1}, Malcolm Kitchen \affil{1}, Neill E. Bowler \affil{2}, Edmund K. Stone \affil{2} }

 \address{
\affilnum{1}Department of Physics and Astronomy, University of Exeter, Exeter, United Kingdom\\
\affilnum{2}Met Office, Exeter, United Kingdom}


 \corraddr{Ollie Lewis (now at MetDesk Ltd.): oliverlewis079@gmail.com}

\begin{abstract}
The extreme variability of humidity in the lower atmosphere over short spatial and temporal scales presents an enormous challenge for existing observing systems. Variations in humidity can be detected due to changes in the refractive properties of the atmosphere, which influence the propagation of electromagnetic radiation. We demonstrate the ability to retrieve high-resolution refractivity gradient measurements using an interferometer to measure the refraction of the Automatic Dependent Surveillance-Broadcast (ADS-B) radio transmission routinely broadcast by commercial aircraft. The ADS-B interferometry (ADSBi) technique is sensitive to small-scale spatial and temporal variability in the refractivity gradient due to changes in humidity. We demonstrate how the retrieval of refractivity profiles is possible exploiting the differences in the observed refraction of radio transmissions broadcast by aircraft at different distances and altitudes.
\end{abstract}


\maketitle

\section{Introduction} 

Automatic Dependent Surveillance-Broadcast (ADS-B) interferometry (ADSBi) is a novel remote sensing technique that has the potential to provide high-resolution refractivity information. The method aims to address a gap in accurate, high-resolution refractivity observations in the lowest few kilometres of the atmosphere above the surface \citep{geer2017growing,leuenberger2020improving}. Variability in refractivity in the lower atmosphere is dominated by variations in water vapour, hence refractivity measurements can provide important humidity information for use in numerical weather prediction (NWP) models. There is increasing interest in novel and opportunistic remote sensing technologies to gather high-volume, low-cost atmospheric data. The ADSBi technique acts as an extension of existing systems used for gathering aircraft-derived observations. Meteorological parameters such as temperature \citep{de2011high, stone2015introducing, mirza2019towards} and wind \citep{de2013use, mirza2016comparison} data have already been derived using ADS-B and Mode-Selective enhanced surveillance (Mode-S EHS) radio transmissions from commercial aircraft. In regions with high air traffic such as Europe, the impact of assimilating wind observations has already been realised by the United Kingdom (UK) Met Office \citep{stone2016network, stone2018comparison, li2021impact} and the Royal Netherlands Meteorological Institute (KNMI) \citep{de2012assimilation, de2013use}.

The ADSBi technique was first introduced by \citet{lewis2023new} along with initial sensitivity tests using synthetic data. The potential to adapt the observation operator used in Global Navigation Satellite System (GNSS) radio occultation was described in \citet{lewis2023new2}. Initial refractivity retrievals using real observational data and a novel adjoint model were explored in \citet{lewis2024refractivity} and \citet{lewis2025retrieving}. The initial refractivity retrievals were impacted by significant observational noise, with the principal source of uncertainty being foreground reflections interfering with the direct ADS-B signal. This paper presents, for the first time, results from the analysis of data that shows good agreement with contemporary radiosonde profiles. The potential to retrieve useful refractivity information is demonstrated. 

In section 2 we present a brief description of the prototype ADS-B interferometer and the set up used in the latest field experiment. In section 3, the observation operator and adjoint model used to retrieve refractivity profiles is described. In section 4, the experimental results are presented alongside a description of the prevailing meteorological conditions.

\section{ADSBi} 

The angle-of-arrival (AoA) of the ADS-B signal is the received elevation angle (angle measured upwards from the horizontal plane) and is determined through the use of a vertically-orientated, two-element interferometer. Atmospheric refraction causes radio signals to deflect and (usually) bend towards the surface, resulting in the apparent elevation of the aircraft to appear greater than its true elevation above the horizon \citep{lewis2023new}. The AoA is determined by measuring the phase difference across the elements of the interferometer (for positive elevations, the signal is received at the upper element before the lower element). The phase difference is given by 

\begin{equation}
 \phi = 2\pi\frac{B\text{sin}(\beta)}{\lambda},
 \label{eq:phase_AoA}
\end{equation}

\noindent where $B$ is the magnitude of the separation vector between the interferometer elements, $\beta$ is the observed AoA and $\lambda$ is the wavelength of the ADS-B signal ($\lambda \approx 0.275$ m for a 1090 MHz signal ).

\begin{figure}
\includegraphics[width=0.5\textwidth]{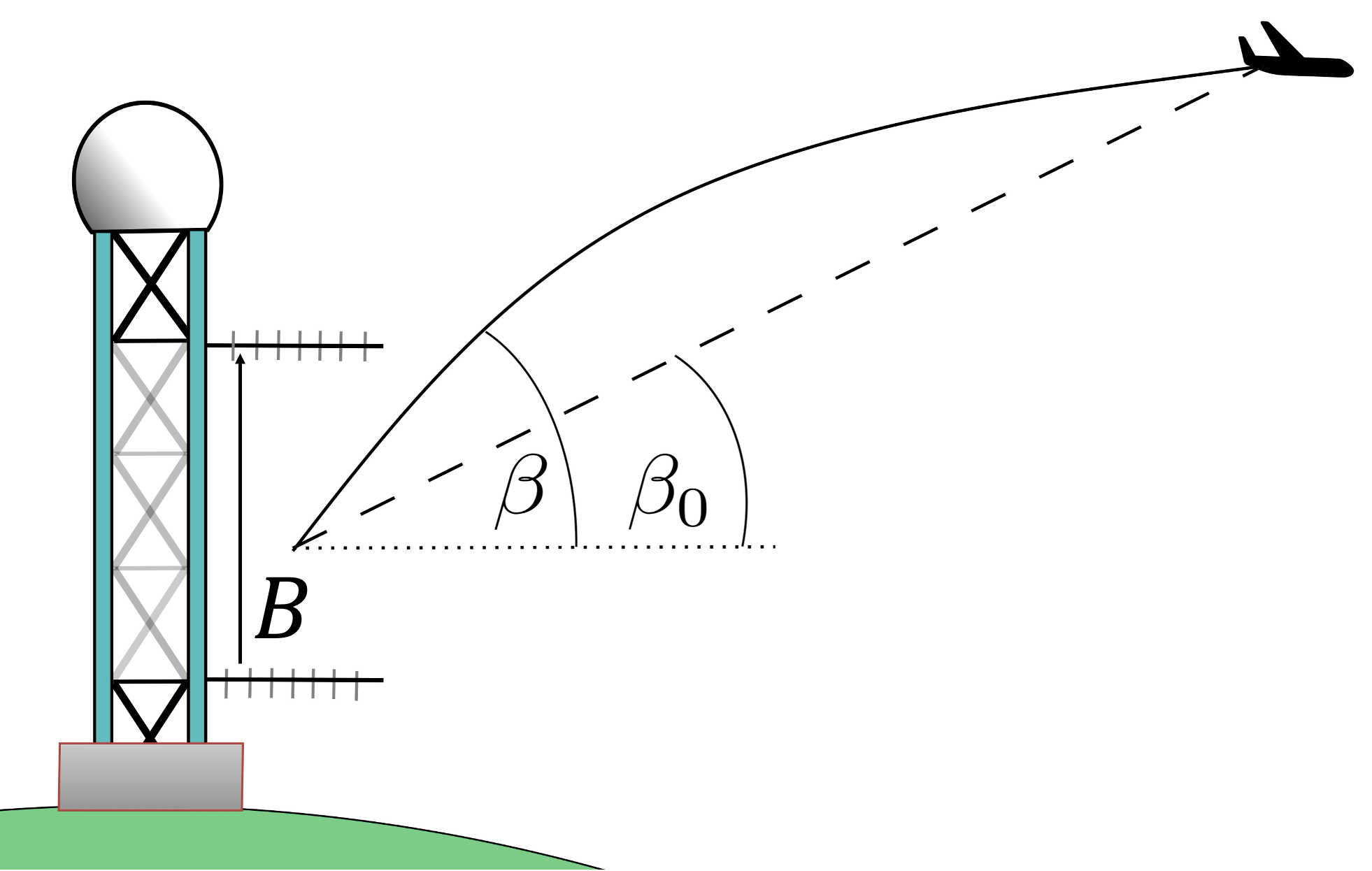}
 \caption{Geometry of the ADSBi observing system. The refracted path of the ADS-B radio signal is given by the solid line. The refracted angle is defined as the difference between the observed AoA ($\beta$) and the LoS AoA ($\beta_0$). The vertically-orientated baseline vector is given by $\boldsymbol{B}$ (adapted with permission from \citet{lewis2024refractivity} ) (© 2024 IEEE).}
 \label{fig:adsb_geometry}
\end{figure}

\noindent The measured AoA of the ADS-B signal can be combined with the known position of the aircraft (information contained within the ADS-B message) to infer the total bending of the signal due to refraction. The interferometrically-derived observed AoA is dependent on the refracted ray path of the ADS-B radio signal. The line-of-sight (LoS) AoA, $\beta_0$, is straight line AoA determined from the reported aircraft position. The difference between the observed and LoS AoA is given by the refracted angle (geometry shown in Fig. \ref{fig:adsb_geometry}), which can be written as

\begin{equation}
 \Delta \beta = \beta - \beta_0.
\end{equation}

\noindent The refracted angle is a function of the accumulated bending along the trajectory of the ray path and is highly sensitive to variations in atmospheric moisture content, particularly in the lower atmosphere \citep{sokolovskiy2001use}. 

\section{Methods} 

\subsection{Observation operator}

\noindent The observation operator allows observations to be estimated given an atmospheric state $\boldsymbol{x}$. The mapping between state and observation space can be represented as

\begin{equation}
 \boldsymbol{y} = \mathcal{H}[\boldsymbol{x}],
\end{equation}

\noindent where $\mathcal{H}$ is a (potentially non-linear) model that transforms the state vector into an equivalent vector of simulated observations $\boldsymbol{y}$. There are two ``observable" quantities obtained using the ADSBi technique: the interferometrically-measured AoA of the signal and the reported height of the aircraft (not directly measured, but extracted from the ADS-B transmission). The observed AoA is challenging to simulate, as it requires tracing a ray from the aircraft position to the receiver position. Two issues arise here: the emission angle of the signal is unknown, and the ray may not necessarily land at the receiver position when traced through a first-guess atmosphere. We instead choose the reported height of the aircraft as the ``observable" to simulate. The reported aircraft height is simulated in a time-reversed frame: the observed AoA is used as the initial direction of the ray, which is then traced out to the reported distance of the aircraft. The height of the ray at this distance can then be compared to the reported height of the aircraft extracted from the ADS-B transmission. The observation operator that maps the state to a simulated ADS-B refraction observation is given by the second-order differential equation (SODE) ray tracer described by \citet{zeng2014radar}. It can be written as two coupled first-order differential equations given by

\begin{equation}
 \frac{\text{d}h}{\text{d}r} = u,
\end{equation}

\begin{equation}
 \frac{\text{d}u}{\text{d}r} = (1-u^2)\bigg(\frac{\text{d}\tilde{\eta}}{\text{d}h} + \frac{1}{a+h}\bigg),
\end{equation}

\noindent where $h$, $u$ and $r$ are the height, direction and along-beam range of the ray respectively. The radius of curvature of the surface of the Earth and the natural logarithm of the refractive index are given by $a$ and $\tilde{\eta}$ respectively. The refractive index of the atmosphere is dependent on the temperature, $T$ [K], pressure, $p$ [hPa] and the partial pressure of water vapour, $e$ [hPa], and can be written as \citep{smith1953constants} 

\begin{equation}
 \eta = 1 + 10^{-6}\left(k_1\frac{p}{T} + k_2\frac{e}{T^2}\right),
\end{equation}

\noindent where $k_1 = 77.6$ K hPa$^{-1}$ and $k_2 = 3.73\times 10^5$ K$^2$hPa$^{-1}$ are empirically derived constants. Other formulae exist for the calculation of refractivity (e.g. \citet{aparicio2011evaluation, healy2011refractivity} ) but the simple form is still commonly used. Generally, the refractive index exceeds unity by only a few hundred parts-per-million (ppm), therefore refractivity is more frequently used to describe variations in the refractive index. Refractivity, $N$, is defined as

\begin{equation}
 N = (n-1)10^6.
\end{equation}

\noindent The total refractivity can be decomposed into a ``dry" (hydrostatic), $N_\text{dry}$, and ``wet", $N_\text{wet}$, component, where

\begin{equation}
 N_\text{dry} = k_1 \frac{p}{T},
\end{equation}

\begin{equation}
 N_\text{wet} = k_2\frac{e}{T^2}.
 \label{eq:wet_refrac}
\end{equation}

\noindent Adopting a similar optimisation approach as described in \citet{lewis2025retrieving}, the target quantity in ADS-B interferometry is the end point altitude of the traced ray from the observer to the distance of the reporting aircraft. The wet refractivity, $N_\text{wet}$, is linearly proportional to the partial pressure of water vapour at a given temperature, making it a direct and physically interpretable proxy for atmospheric humidity content.

\subsection{Variational retrievals }
 \label{subsec:var_retrievals}

The maximum likelihood state can be determined by minimising the cost function

\begin{equation}
\begin{split}
 \mathcal{J}(\boldsymbol{x}) = \frac{1}{2}(\boldsymbol{x} - \boldsymbol{x}_{\text{b}})^{\text{T}}\textbf{B}^{-1}(\boldsymbol{x} - \boldsymbol{x}_{\text{b}}) \\
 +\frac{1}{2}(\boldsymbol{y}_\text{o}-\mathcal{H}[\boldsymbol{x}])^{\text{T}}\textbf{R}^{-1}(\boldsymbol{y}_\text{o}-\mathcal{H}[\boldsymbol{x}]),
\end{split}
\label{1dvar_eq}
\end{equation}

\noindent with respect to the state vector $\boldsymbol{x}$. The background state vector and observations are given by $\boldsymbol{x}_\text{b}$ and $y_\text{o}$ respectively. The background and observation error covariance matrices are given by \textbf{B} and \textbf{R} respectively. The optimum state can be determined using a gradient descent algorithm, which requires the gradient of $\mathcal{J}$ with respect to the state vector $\boldsymbol{x}$. The gradient is given by

\begin{equation}
 \frac{\text{d}\mathcal{J}(\boldsymbol{x})}{\text{d}\boldsymbol{x}} = \textbf{B}^{-1}(\boldsymbol{x} - \boldsymbol{x}_{\text{b}}) - \bigg(\frac{\text{d}\mathcal{H}[\boldsymbol{x}]}{\text{d}\boldsymbol{x}}\bigg)^\text{T}\textbf{R}^{-1}(\boldsymbol{y}_\text{o}-\mathcal{H}[\boldsymbol{x}]).
\end{equation}

\noindent Introducing an additional constraint on the ray emission direction, the cost function described by \eqref{1dvar_eq} for a single aircraft radio transmission can be written as

\begin{equation}
\begin{split}
 j(\tilde{\boldsymbol{\eta}}) = \frac{1}{2}(\tilde{\boldsymbol{\eta}} - \tilde{\boldsymbol{\eta}}_{\text{b}})^{\text{T}}\textbf{B}^{-1}(\tilde{\boldsymbol{\eta}} - \tilde{\boldsymbol{\eta}}_{\text{b}}) \\
 + \frac{1}{2}\frac{(h_\text{o}-h[\tilde{\boldsymbol{\eta}}])^2}{\sigma_h^2} +
 \frac{1}{2}\frac{(u[\tilde{\boldsymbol{\eta}}_{\text{b}}]-u[\tilde{\boldsymbol{\eta}}])^2}{\sigma_u^2}, 
\end{split}
\label{eq:1dvar_eq_eta}
\end{equation}

\noindent where $\sigma_h$ and $\sigma_u$ are the observation error uncertainties associated with the end height and direction for a single ray. The observation uncertainties are assumed to be uncorrelated, therefore off-diagonal elements of \textbf{R} are assumed to be zero (observational uncertainties may be correlated due to foreground reflections, although determining the exact nature of the observation error covariance matrix is beyond the scope of this study). Since the emission direction of the radio transmission from the aircraft is unknown, it is instead estimated from the background state as

\begin{equation}
 u[\tilde{\boldsymbol{\eta}}_{\text{b}}] = \text{sin}\bigg(\text{cos}^{-1}\bigg(\frac{\eta_0(a+h_0)\text{cos}(\beta)}{\eta_{\text{T}}(a+h_{\text{T}})}\bigg)\bigg),
 \label{eq:bg_u}
\end{equation}

\noindent where $\eta_0$ and $h_0$ are the refractive index and height of the ray at the location of the receiver. The refractive index, $\eta_{\text{T}}$, at the height of the aircraft, $h_{\text{T}}$, is determined from the background state $\boldsymbol{\eta}_{\text{b}}$. The cost function for a single ray defined by \eqref{1dvar_eq_eta} can be augmented with the known physics of the forward operator in a Lagrange multiplier approach. The Lagrangian can be written in terms of the cost function as

\begin{equation}
\begin{split}
 \mathcal{L} = j
 - \int_0^{r_\text{f}}\psi(\dot{h}-u)\text{d}r \\
 - \int_0^{r_\text{f}}\mu\bigg(\dot{u}-(1-u^2)\bigg(\frac{\text{d}\tilde{\eta}}{\text{d}h}+\frac{1}{a+h}\bigg)\bigg)\text{d}r,
\end{split}
\label{1dvar_eq_eta}
\end{equation}

\noindent where the final two terms constrain the solution. The adjoint variables associated with $h$ and $u$ are given by $\psi$ and $\mu$ respectively. The adjoint variables are initialised as

\begin{equation}
 \psi = \frac{\partial j}{\partial h} = \frac{(h_o - h[\tilde{\boldsymbol{\eta}}])}{\sigma_h^2},
\end{equation}

\begin{equation}
 \mu = \frac{\partial j}{\partial u} = \frac{(\tilde{u}_o - u[\tilde{\boldsymbol{\eta}}])}{\sigma_u^2}.
\end{equation}

The uncertainties in $h$ and $u$ are difficult to determine, since the ray path is a non-linear, integrated quantity that depends on the uncertainty in the observed AoA and the aircraft altitude reporting resolution. For this study, $\sigma_h = 0.1$ km and $\sigma_u=0.1$ are used respectively. More accurate values are a topic of further investigation. The variation of the cost function for a single ray is given by

\begin{equation}
 \delta \mathcal{L} = \delta\tilde{\boldsymbol{\eta}}\textbf{B}^{-1}(\tilde{\boldsymbol{\eta}} - \tilde{\boldsymbol{\eta}}_{\text{b}}) + \int_0^{r_\text{f}}\mu(1-u^2)\bigg(\frac{\text{d}(\delta\tilde{\eta})}{\text{d}h}\bigg)\text{d}r,
\label{1dvar_eq_eta_var}
\end{equation}

\noindent representing the interpolated (log) refractive index as $\tilde{\eta} = [w_0, w_1]^\text{T}[\text{ln}(\eta_0), \text{ln}(\eta_1)]$, the gradients of the Lagrangian with respect to the refractive index values $\eta_0$ and $\eta_1$ are given by

\begin{equation}
 \frac{\text{d}\mathcal{L}}{\text{d}(\text{ln}(\eta_0))} = w_0 \hat{\textbf{B}} _0 + \int_0^{r_\text{f}}\mu(1-u^2)\bigg(\frac{\text{d}w_0}{\text{d}h}\bigg)\text{d}r,
 \label{eq:grad_1}
 \end{equation}

\begin{equation}
 \frac{\text{d}\mathcal{L}}{\text{d}(\text{ln}(\eta_1))} = w_1 \hat{\textbf{B}} _1 + \int_0^{r_\text{f}}\mu(1-u^2)\bigg(\frac{\text{d}w_1}{\text{d}h}\bigg)\text{d}r,
 \label{eq:grad_2}
 \end{equation}

\noindent where $\hat{\textbf{B}} = \textbf{B}^{-1}(\tilde{\boldsymbol{\eta}} - \tilde{\boldsymbol{\eta}}_{\text{b}})$. The background state constrains the solution and prevents the retrieved refractive index deviating too strongly from the initial state.

The gradients described by \eqref{eq:grad_1} and \eqref{eq:grad_2} are used to update the best estimate of the refractive index, $\boldsymbol{\eta}$. However, attempting to retrieve a vertical refractivity profile using a single AoA measurement is an ill-posed problem, as there are numerous ray paths that satisfy a given AoA and ray end point. Instead, numerous ADS-B transmissions from multiple different aircraft in different locations are used to constrain the retrieved refractivity profile. Therefore, the total cost function, $\mathcal{J}$ is the sum of all the cost functions associated with individual AoA measurements, written as

\begin{equation}
 \mathcal{J} = \sum^{N_\text{B}}_i j_i,
 \label{eq:total_cost}
\end{equation}

\noindent where $i$ is an index and $N_\text{B}$ is the total number of ADS-B AoA measurements used in the retrieval. A more detailed overview of the optimisation process is described in \citet{lewis2025retrieving}. The criteria used to stop the minimisation were:

\begin{itemize}
 \item $\frac{\mathcal{J}_{i+1}-\mathcal{J}_i}{\mathcal{J}_0} < 0.01$, where $\mathcal{J}_0$ is the initial value of the cost function, or;
 \item number of iterations exceeds ten.
\end{itemize}

 With each iteration step, the initial refractivity profile is perturbed sequentially by the gradients given by equations \eqref{eq:grad_1} and \eqref{eq:grad_2} computed using each individual ray. The cost function for each iteration step is then computed by summing the individual cost functions for each ray, as given by equation \eqref{eq:total_cost}. A more detailed description of the method is given by \citet{lewis2025retrieving}, and more generally for optimisation problems of this kind by \citet{teh2022adjoint}.

 \section{Analysis and results}

\begin{figure}[b]
\includegraphics[width=0.5\textwidth]{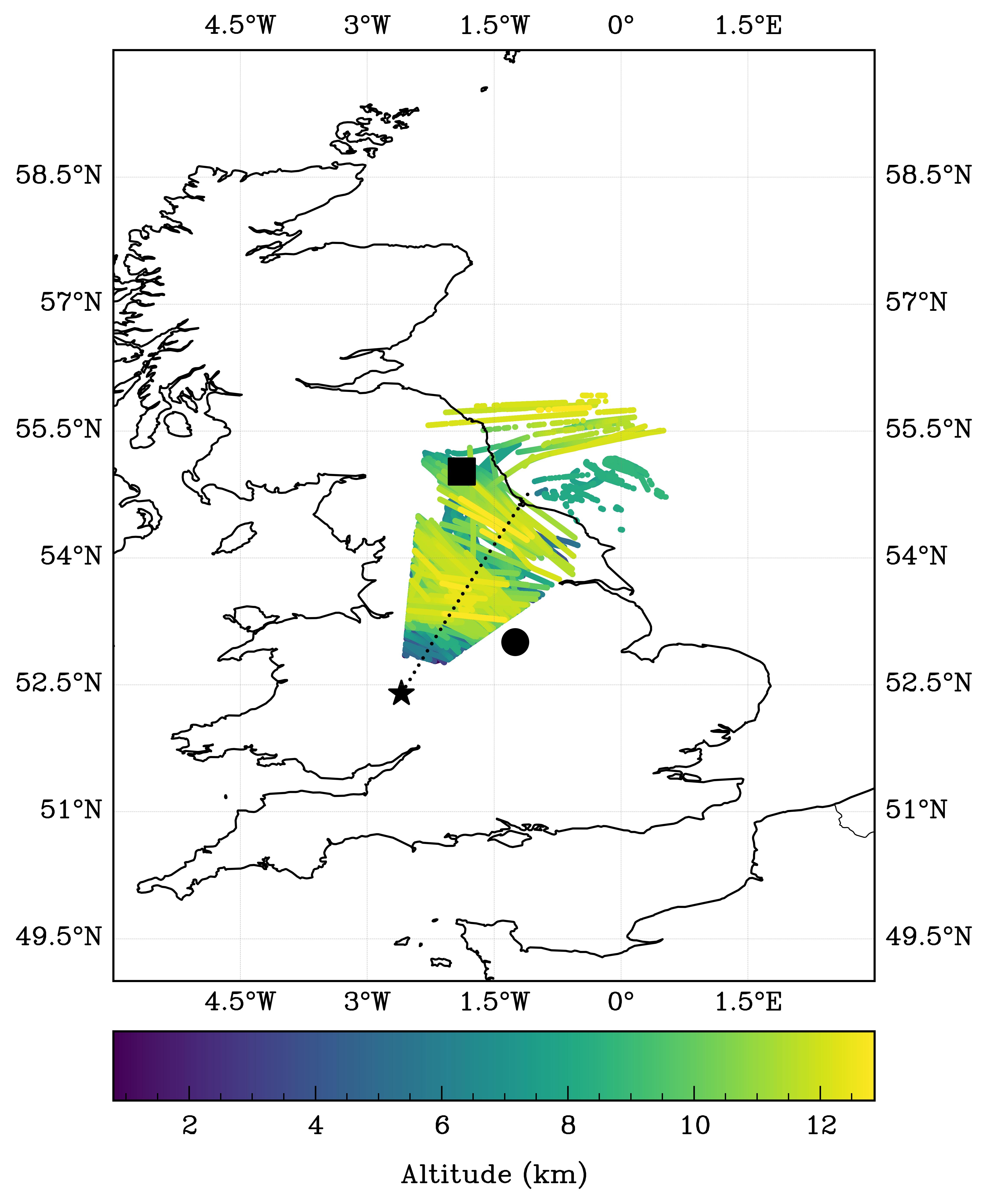}
 \caption{Distribution of the reported aircraft positions (latitude and longitude) extracted from the received ADS-B transmissions received throughout the observation period from 09:00 UTC to 15:30 UTC. The location of the ADS-B interferometer is indicated by the black star. The locations of the Watnall and Albemarle radiosonde launch station are indicated by the black circle and square respectively. Each reported position is colour-coded by the reported altitude. The dashed line shows the coordinates of the representative ray path from which the refractivity profile was extracted from the UKV model. }
 \label{fig:map}
\end{figure}

Approximately 5 million ADS-B transmissions were received within an AoA of $-0.6\degree$ and $3.0\degree$ between 09:00 UTC and 15:30 UTC on the 4 September 2024. Only ADS-B transmissions beyond 50 km distance from the receiver were used (uncertainty in the ADS-B reported position closer than this result in large errors in the LoS AoA, reducing the accuracy of the refracted angle measurement). The geographical distribution of ADS-B signals received by the interferometer is shown in Fig. \ref{fig:map}.

\subsection{Meteorological conditions}

\begin{figure}
 \centering
 \begin{subfigure}[b]{0.45\textwidth}
 \centering
 \includegraphics[width=\textwidth]{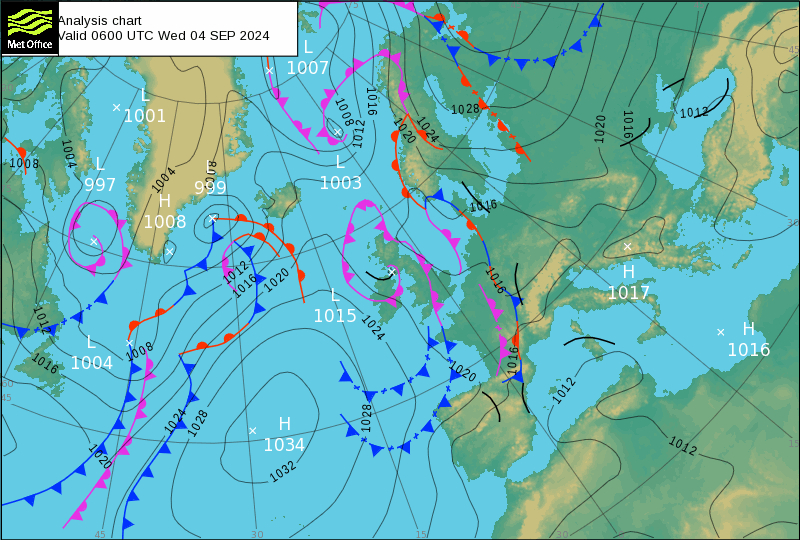}
 \caption{}
 \label{fig:y equals x}
 \end{subfigure}
 \hfill
 \begin{subfigure}[b]{0.45\textwidth}
 \centering
 \includegraphics[width=\textwidth]{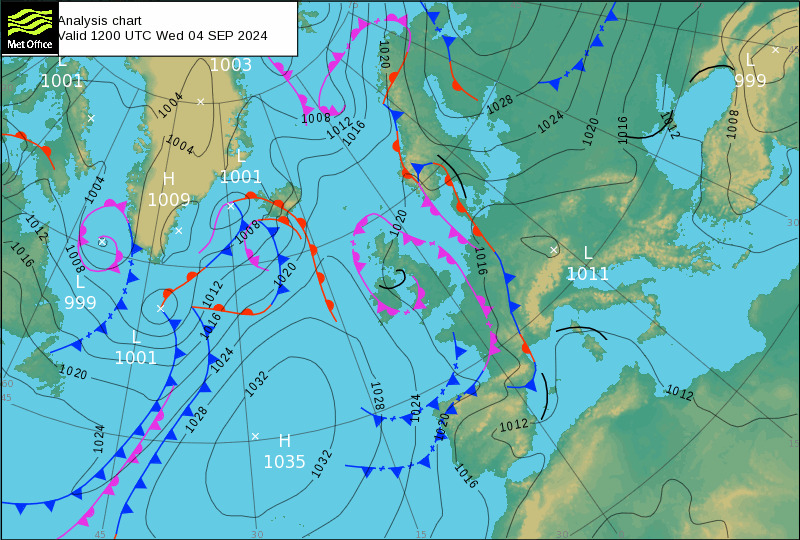}
 \caption{}
 \label{fig:three sin x}
 \end{subfigure}
 \hfill
 \begin{subfigure}[b]{0.45\textwidth}
 \centering
 \includegraphics[width=\textwidth]{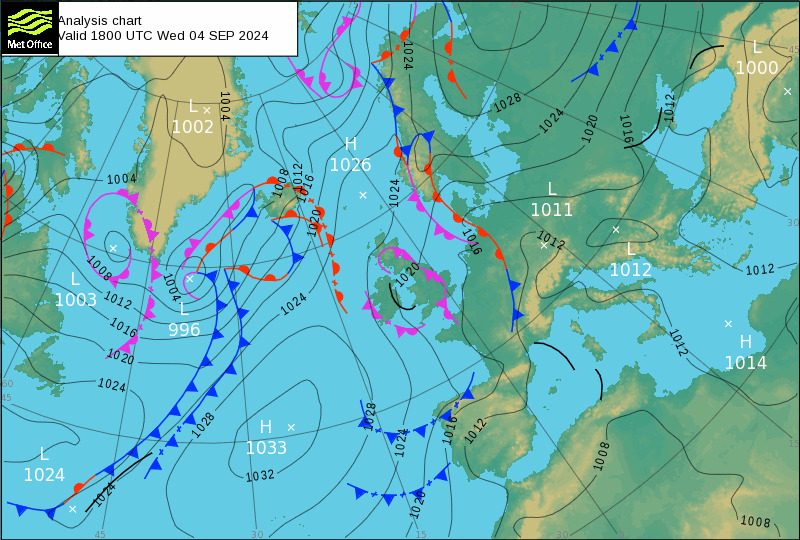}
 \caption{}
 \label{fig:five over x}
 \end{subfigure}
 \caption{Synoptic charts for the 4 September 2024 at (a) 06:00 UTC, (b) 12:00 UTC and (c) 18:00 UTC (© British Crown copyright 2026, the Met Office).}
 \label{fig:synoptic}
\end{figure}

\begin{figure}
\includegraphics[width=0.5\textwidth]{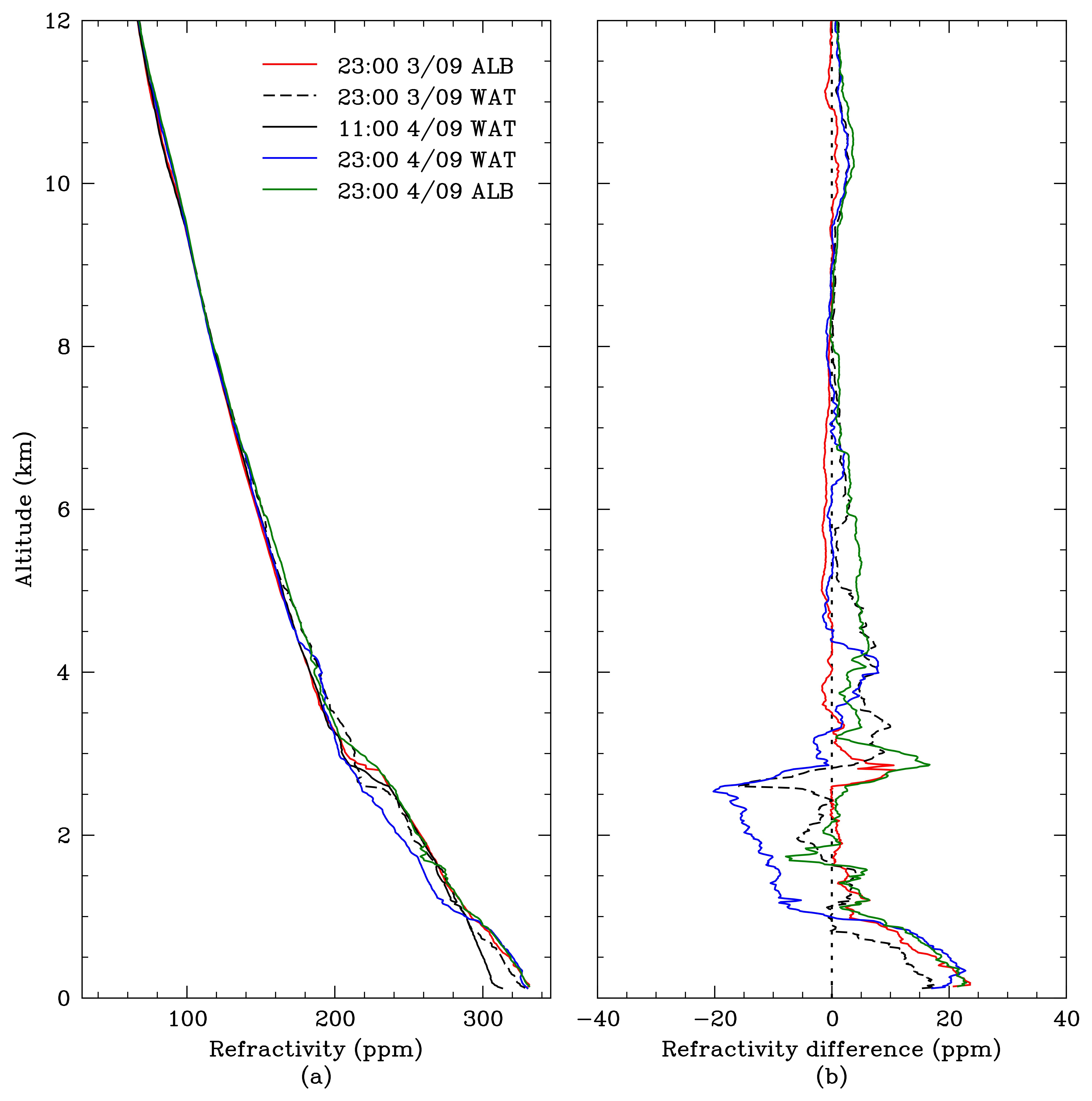}
 \caption{(a) Vertical refractivity profiles obtained using data from various Watnall (WAT) and Albemarle (ALB) radiosonde launches for the 3 and 4 September 2024. (b) Difference between the vertical refractivity profiles and the 11:00 UTC refractivity profile obtained using data from the Watnall radiosonde launch on the 4 September 2024 (solid black in (a)).}
 \label{fig:sondes}

\end{figure}

The synoptic pattern across the UK during the observation period on the 4 September 2024 is shown in Fig. \ref{fig:synoptic}. At 06:00 UTC an occluded front was present across central parts of the UK, which by 12:00 UTC moved eastwards, clearing into the North Sea. The occluded front then travelled across the observation region again between 12:00 UTC and 18:00 UTC. Winds were relatively light during this period, with slack isobars across the UK. Figure \ref{fig:sondes} shows the vertical refractivity profiles derived from data using various radiosonde launches at Watnall and Albemarle. Figure \ref{fig:sondes}b shows the variability in the vertical refractivity profiles relative to the reference 11:00 UTC 4 September 2024 (04/09) radiosonde profile from Watnall. All profiles show a significant increase in refractivity relative to the 11:00 UTC 04/09 Watnall profile within the first kilometre above the surface. There is generally a decrease in refractivity above this altitude for the other profiles relative to the reference profile, with the 23:00 UTC 04/09 Watnall profile showing the largest decrease.






\subsection{Refracted angle observations}

\begin{figure*}
 \includegraphics[width=\textwidth]{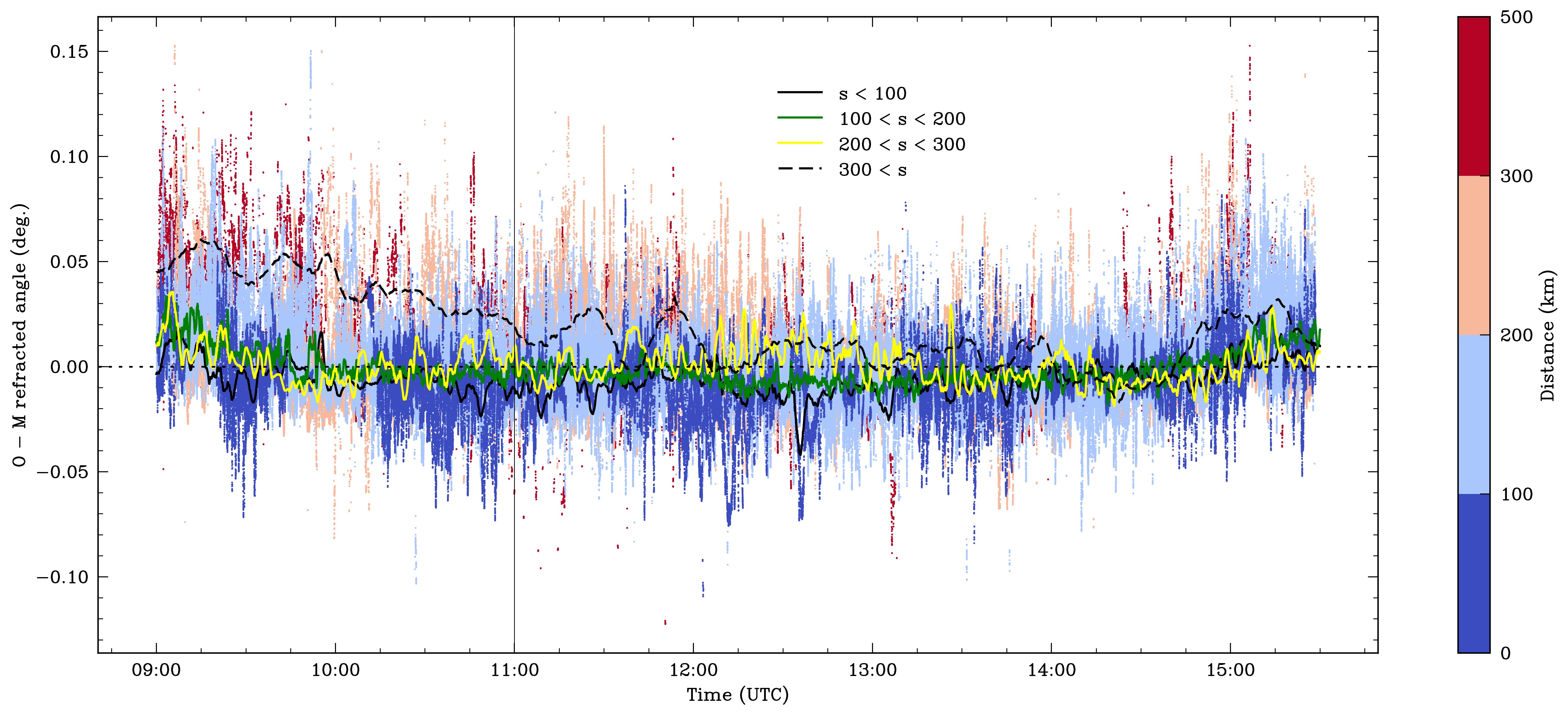}
 \caption{Difference between the observed (O) and modelled (M) refracted angle for the ADS-B transmissions received between an observed AoA of $0.0\degree$ and $3.0\degree$ as a function of time for 04/09. Each point is an individual received ADS-B transmission, colour-coded by the reported distance. The mean O - M refracted angle for various distance ($s$, [km]) bins are indicated the dashed and solid curved lines. The 11:00 UTC Watnall radiosonde launch is indicated by the solid vertical line. }
 \label{fig:obs_minus_model_time}
\end{figure*}

The difference between the observed and modelled refracted angle as a function of time is shown in Fig. \ref{fig:obs_minus_model_time}, where each point is an individual received ADS-B transmission. The reference vertical refractivity profile used to model the refraction was obtained using data from the 11:00 UTC 04/09 Watnall radiosonde launch. This was the radiosonde launch closest in time to the observation period. The observed minus model plot shows temporal variability, with refraction measurements early in the observation period showing the largest deviation from the reference model. There is also significant spatial variability, particularly within the first hour of observations. Aircraft with reported distances greater than 300 km show significantly stronger refraction than nearby aircraft early on in the observation period, before the deviation from the reference refraction model decreases for all reported distances later. The mean refracted angle difference shows significant spread, particularly at earlier times, perhaps indicating a complex refractivity field. The synoptic charts and refractivity profiles shown in Fig. \ref{fig:synoptic} and \ref{fig:sondes} respectively show that an occluded front passed through the observation region during the morning, with an associated decrease in refractivity within the first kilometre altitude above the surface between the 23:00 UTC 03/09 and 11:00 UTC 04/09 Watnall radiosonde profiles (Fig. \ref{fig:sondes}). This drop in surface refractivity was in agreement with in-situ measurements taken at the interferometer location. The passage of the occluded front perhaps resulted in a decrease in the vertical refractivity gradient, weakening the observed refraction. 

 \subsection{Refractivity retrievals}

\begin{figure*}
\includegraphics[width=\textwidth]{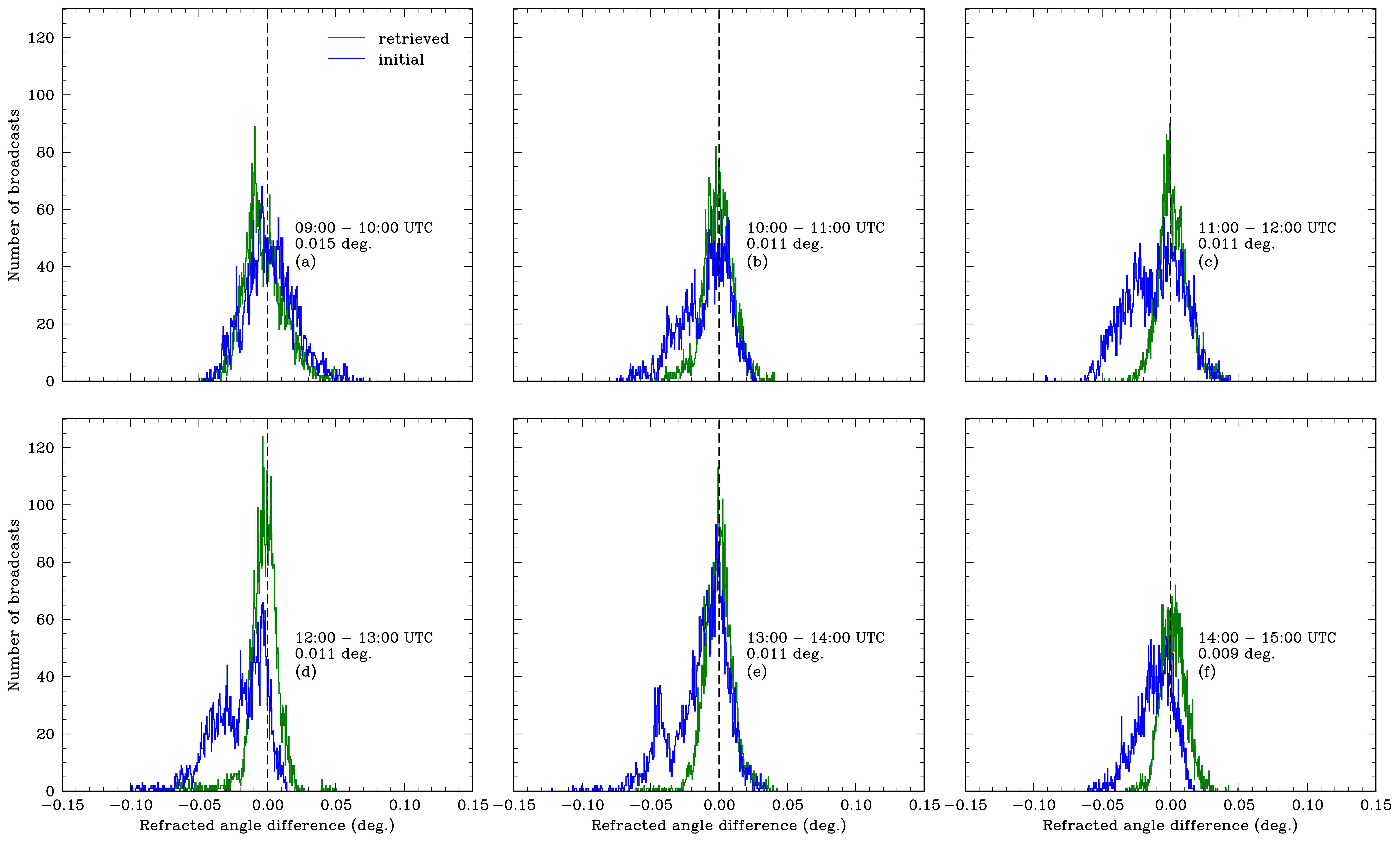}
 \caption{The distribution in the refracted angle difference for all observations using the initial (blue) and retrieved (green) refractivity profiles. The standard deviation in the retrieved refracted angle difference is shown, with the largest standard deviation being $0.015\degree$ during the 09:00 to 10:00 UTC observation period. }
 \label{fig:retrieval_4sep_hist}
\end{figure*}

 Refractivity retrievals were performed for the wet component, $N_\text{wet}$, rather than total refractivity, since the dry component, $N_\text{dry}$, is well-constrained by the known pressure and temperature structure of the atmosphere and contributes little uncertainty to the retrieval. The wet refractivity is linearly proportional to the partial pressure of water vapour at a given temperature, making it a direct and physically interpretable proxy for atmospheric humidity content, and the primary source of variability in refractivity in the lower troposphere. The one-dimensional vertical wet refractivity profile was retrieved using the 04/09 observations in one-hour intervals. The retrievals used 5000 ADS-B transmission s received within an observed AoA of $0.0\degree$ to $2.0\degree$. As described in Section \ref{subsec:var_retrievals}, retrievals using a single transmission are ill-posed, so the retrieval algorithm is run sequentially for the 5000 rays within a time window to return a refractivity profile that is best fit for the observed refraction of all rays within the time window. The spherical symmetry assumption is reasonable for ray tracing positive elevation rays in most cases, since the vertical variation in refractivity far dominates over horizontal variations. The azimuthal width of the observation sector was restricted to the central $20\degree$ for the retrievals. The difference between the observed and the simulated LoS AoA using the initial and retrieved refractivity profiles are shown in Fig. \ref{fig:retrieval_4sep_hist}. The retrieved refractivity profile results in a shift in the mean difference towards $0\degree$ and a reduction in variance, reflecting internal consistency and indicating the retrieval method effectively reduces the difference between the simulated and observed refracted angle via the optimisation of the vertical refractivity profile.

\begin{figure*}
 \includegraphics[width=\textwidth]{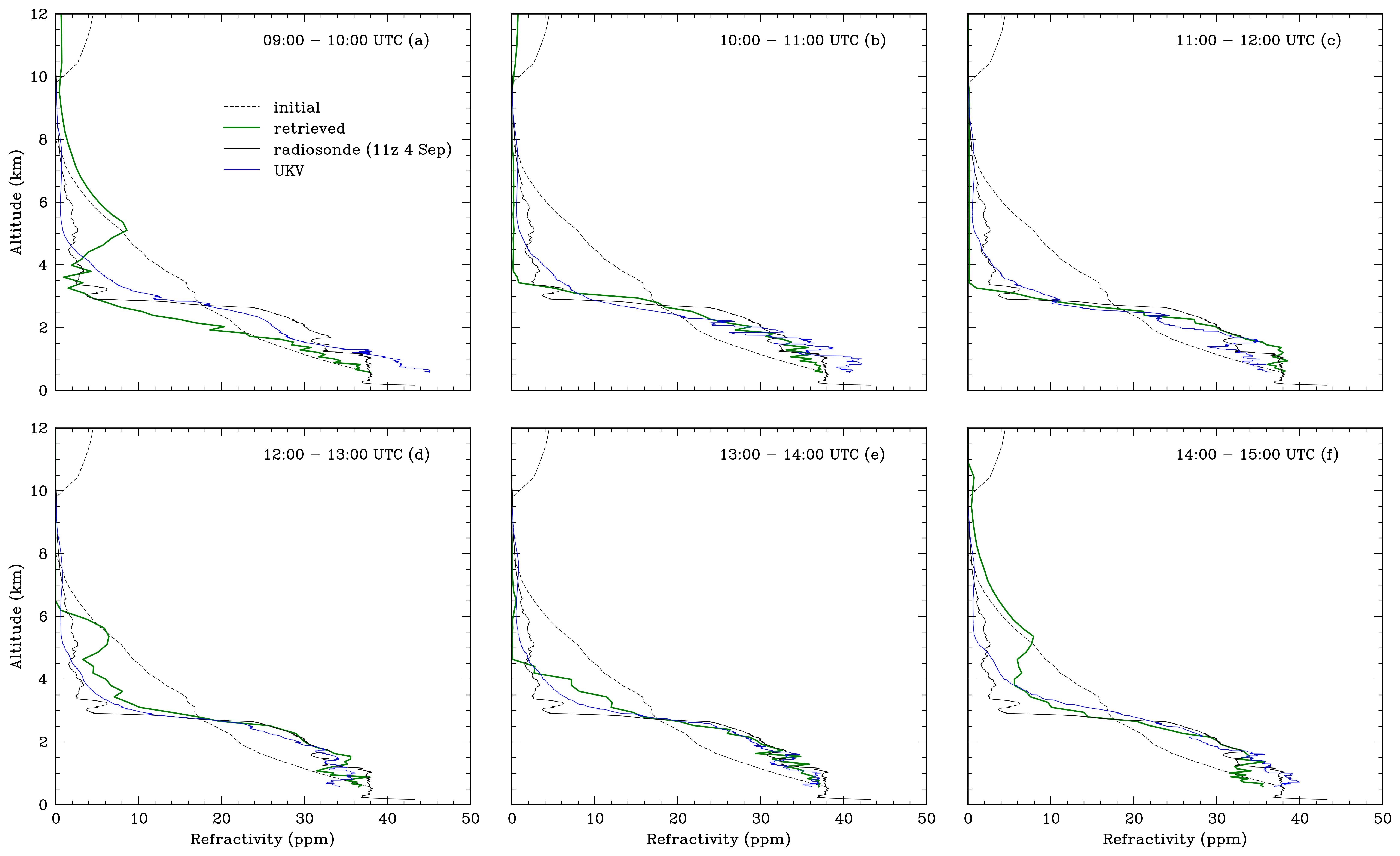}
 \caption{The retrieved (green) and initial (dashed) vertical wet refractivity profiles for each of the six one-hour observation windows (a – f) from 09:00 to 15:00 UTC on 4 September 2024. For each time window, the high-resolution UKV model refractivity profile extracted along a representative ray is shown (solid blue line), providing an hourly NWP comparison throughout the observation period. The 11:00 UTC radiosonde profile from Watnall is shown for reference ( solid black line ). The refractivity at the receiver altitude was fixed to the interpolated value from in-situ measurements at the receiver location.}
 \label{fig:retrieval_4sep}
\end{figure*}

Refractivity profiles extracted from the Met Office's United Kingdom Variable-resolution (UKV) model \citep{tang2013benefits} along a representative ray (the latitude and longitude positions are shown by the dotted line in Fig. \ref{fig:map}) for each of the six one-hour retrieval windows and are shown alongside the retrievals in Fig. \ref{fig:retrieval_4sep} (solid blue line), providing an hourly high-resolution NWP comparison throughout the observation period.
 The retrievals were also assessed against a representative radiosonde sounding at 11:00 UTC from Watnall. Though the Watnall site is outside the observation region shown in Fig. \ref{fig:map}, it was the closest available source of in-situ observations and was likely the most representative profile to assess against, particularly for the lowest few kilometres above the surface (the region of the atmosphere that the ADSBi technique is most sensitive to). To assess the humidity information content of the observations, the retrievals were performed by assigning an altitude-dependent weight to the background constraint introduced by \eqref{eq:bg_u}. The error covariance $\sigma_u$ decreased exponentially with altitude, constraining the emission direction of the ADS-B transmission at high altitudes (where the refractivity is dominated by the well-known dry density structure). The background contribution to the cost function described in \eqref{eq:1dvar_eq_eta} was set to zero (i.e. $\sigma_\text{b} \rightarrow \infty$). The wet refractivity retrievals are shown in Fig. \ref{fig:retrieval_4sep}. The vertical refractivity profile was initialised as a simple exponential with the surface refractivity determined using in-situ measurements of temperature and humidity at the receiver location. At each iteration during the variational minimisation procedure the refractivity was clamped between the dry and saturated refractivity profiles (i.e. the relative humidity was fixed between a value of $0\%$ and $100\%$). 

 The retrieved wet refractivity profiles show significant variability compared to the reference radiosonde and UKV profiles, with the 09:00 UTC retrieval showing the largest deviation. This coincides with the significant distance-dependent departure from the reference 11:00 UTC 04/09 Watnall refraction model shown in Fig. \ref{fig:obs_minus_model_time}. The retrievals using later times show better agreement with the reference radiosonde and UKV profiles, particularly when capturing the steep gradient in wet refractivity at $\sim 3$ km altitude. Despite this, the retrievals are relatively noisy with spurious refractivity structure near the surface and at $\sim 5$ km altitude using the 10:00-11:00 UTC and 14:00-15:00 UTC observational data.

 In this preliminary test, we have modelled the atmosphere using a one-dimensional refractivity profile, which is inadequate to represent the true three-dimensional structure of the refractivity field. Figure \ref{fig:obs_minus_model_time} shows that the refracted angle measurements are sensitive to spatial and temporal variability in the refractivity field, but obviously a one-dimensional retrieval cannot fully exploit this information. Consequently, refractivity variations in distance, transverse to the line-of-sight, and on timescales shorter than the one-hour sample window, will therefore manifest as `noise’ in the minimisation. Figure \ref{fig:sondes}b shows that temporally-coincident wet refractivities from Albemarle and Watnall differ in complex ways by approximately 5-10 ppm, comparable to deviations between the 11:00 UTC 4 September Watnall profile and the temporally-coincident retrieval. 

\section{Discussion}

The time series of refracted angle observations obtained using the ADS-B interferometer show detectable spatial and temporal variability in the refractivity environment over sub-hourly timescales. However, the retrievals are currently noisy, particularly during periods with complex distance-dependent structure in the refraction observations. A number of potential sources of uncertainty were explored in a previous investigation by \citet{lewis2025retrieving}. A potential source of observational uncertainty was the presence of frontal systems, which introduce significant horizontal gradients in refractivity. The ray tracing model developed by \citet{zeng2014radar} assumes that the refractivity exhibits spherical symmetry (i.e. the refractivity field only varies radially). It was expected that this was generally a reasonable assumption when modelling the refraction of rays with a positive AoA. However, an occluded front passed through the observation region between 06:00 and 12:00 UTC (Fig. \ref{fig:synoptic}) which may have introduced more significant horizontal variability. The current validation uses (approximately) vertical radiosonde profiles versus `slant’ refraction-derived profiles (i.e. along the ADS-B signal propagation path). Ultimately, this is a limited approach, and future retrieval techniques must embrace the full three-dimensional information contained in the refracted angle measurements ( Figure \ref{fig:obs_minus_model_time} ). Potentially profitable envisioned development paths include:

\begin{itemize}
 \item development of a two-dimensional adjoint model to capture variations in refractivity with distance;
 \item use of a second, transversely-oriented, ADSBi instrument on another suitably-located radar tower, to measure transverse refractivity gradients that we are currently not directly sensitive to;
 \item ultimately, multiple receivers sampling the same atmospheric volume could be used in a tomographic approach.
\end{itemize}

Adjoint-based retrievals of three-dimensional refractive structures using tomography/embedded sources have been recently demonstrated in other contexts: in turbulent fluids (using tomography) \citep{teh2022adjoint} and in gravitational (general relativistic) bending by dark matter (using embedded galactic sources) \citep{zhao2024single}. Since we have opportunistic access to embedded sources (aircraft), and the future prospect of tomography, the adjoint state technique may be very effective for the ADSBi application in three dimensions.

If the full 3D information can be exploited, an important remaining challenge will be to assemble suitable validation data sets, as other data sources lack comparable spatio-temporal sampling. Quantifying the information content in the ADSBi data may be better approached by assimilation methods, to determine the magnitude and spatio-temporal structure of the resulting forecast impact. Sources of systematic biases in the retrievals include instrumental uncertainties and an accurate measurement of the surface refractivity. The impact of the these uncertainties would be difficult to separate from the effects of a complex refractivity environment on the observations. However, the initial analysis of the initial refractivity retrievals and the time series observations of the refracted angle does not indicate any significant systematic bias. Any instrumental source of uncertainty would likely be far easier to diagnose during periods of calm weather with little variability over short spatial and temporal scales. 
 With longer periods of continuous observations in a variety of atmospheric conditions, it may be possible to properly diagnose the principal sources of systematic noise in the observations.

\section{Conclusions}

The Automatic Dependent Surveillance-Broadcast (ADS-B) interferometry (ADSBi) technique has been developed to measure the refracted angle of radio transmissions broadcast by commercial aircraft. The instrument is highly sensitive to changes in atmospheric conditions over short spatial and temporal scales. The adjoint retrieval model developed by \citet{lewis2025retrieving} has been modified to introduce constraints on the emission angle of the radio transmissions from the broadcasting aircraft at high altitudes (where the impact of water vapour is negligible). Vertical wet refractivity profiles in the lower atmosphere over hourly intervals were retrieved using ADS-B refraction observations obtained on the 4 September 2024. The retrievals show significant variability, particularly at earlier times. However, the profiles show some agreement with the reference radiosonde profiles from the nearby Watnall and Albemarle stations, particularly when capturing the steep refractivity gradient present at $\sim 3$ km altitude. Sources of uncertainty were explored and future work was suggested, such as improvements in the instrument design to minimise multipath contamination and the development of a two-dimensional adjoint model to capture horizontal variations in refractivity.

\section{Acknowledgments}

This work was supported by the Met Office, the University of Exeter and the Harry Otten Foundation. The authors would like to thank Met Office engineers Mike Protts, Liam O'Brien and Andy Wilson for their support with the installation of the interferometer on the Clee Hill weather radar tower. The authors acknowledge the support of other Met Office engineering teams. The radiosonde data used to support this study was obtained from: Met Office (2025) Watnall station high resolution radiosonde data v2. NERC EDS Centre for Environmental Data Analysis (last accessed: 2 September 2025) (\url{https://catalogue.ceda.ac.uk/uuid/7b1c3cc934a840c2b522d847e921c5be}) \citet{MetOffice_2025}.

\section{Competing interests}

The contact author has declared that none of the authors has any competing interests.

\section{Copyright statement}

The works published in this journal are distributed under the Creative Commons Attribution 4.0 License. This license does not affect the Crown copyright work, which is re-usable under the Open Government Licence (OGL). The Creative Commons Attribution 4.0 License and the OGL are interoperable and do not conflict with, reduce or limit each other. © Crown copyright 2026

\bibliographystyle{apalike}
\bibliography{bibliography}

\end{document}